\documentclass[aip,jmp,bmf, sd, rsi, amsmath,amssymb, reprint]{revtex4-1}
\usepackage{graphicx}
\usepackage{bigints}

\begin{document}

\title{Electron transport in quasi-ballistic FETs subjected to a magnetic field}

\author{M. Yelisieiev}%
\email{mykola.eliseev@gmail.com}
\affiliation{Institute of Semiconductor Physics,
National Academy of Sciences of Ukraine, Kyiv 03028, Ukraine;}%

\author{V. A. Kochelap}%
\affiliation{Institute of Semiconductor Physics,
National Academy of Sciences of Ukraine, Kyiv 03028, Ukraine;}%
\affiliation{Institute of High Pressure Physics, Polish Academy of Sciences, Warsaw 01-142, Poland}%

\begin{abstract}
We report the study of quasi-ballistic electron transport in short
FETs subjected to magnetic field. Spatial distributions of
electron concentrations, velocities, Hall currents and voltages in
the short FET channels are determined.  The basic properties of
current-voltage characteristics of quasi-ballistic FETs in magnetic
field are analyzed, among these the kink-like characteristics
of the near-ballistic device. Peculiarities of magnetoresistance of
such FETs are studied for low and high magnetic fields,
and different current regimes. For nonlinear current regimes
we revealed significantly larger magnetoresistance  for the devices
with higher ballisticity. Numerical estimates of studied effects are
presented. We suggest, that the found results contribute
to the physics of short FETs and can be used for developing
nanoscale devices for particular applications.

\end{abstract}

\maketitle

\section{Introduction}

In semiconductor structures and devices with
short (submicron) conductive channels electrons
experience only a few scatterings by defects and
phonons during time of their transit through the devices.
Physics of these ballistic or quasi-ballistic electron transport regimes
is distinctly different from that of essentially dissipative (drift-diffusion)
regimes characteristic for microscale devices.
Examples of new effects emerging in near-ballistic device regimes include
both quantum~\cite{Mitin} and semiclassical phenomena. Among the latter
we may cite velocity overshoot,~\cite{Mitin}
manifestation of nonparabolic energy dispersion,~\cite{Zoric-1}
run-away effect,~\cite{run-away}
excitation of plasmonic resonances,~\cite{pl-resonance-1,pl-resonance-2}
a large effect of lattice vibrations on the ultrahigh frequency
transport in polar materials, \cite{opt-ph-1,opt-ph-2}
 etc. Exploitation of these phenomena may open a way to new classes of
nanoelectronic and optoelectronic devices.

Of nanoscale devices, field effect transistors (FETs) with short active
channels hold an important position because of interesting physics and
numerous areas of
their applications. A great number of papers was dedicated to
experimental and theoretical studies of
these nanoscale devices (see, for example, [\onlinecite{Natori,Lundstrom,Knap}]).
Particularly, in the focus of attention were and remain spatial distributions
and temporal dependencies of electric potentials, charge carriers
(primary theoretical studies) and kinetic characteristics of the
carriers, etc.
Additional information on key characteristics in small sized FETs
can be obtained by using magnetic field, which affects  carrier motion.
Experimental measurements for small sizes FETs in magnetic fields
transverse to electron transport were
reported in papers,~\cite{H-field-1,H-field-2,H-field-3} where the authors have used the theory of dissipative galvanomagnetic transport.
However, in ballistic and quasi-ballistic transport regimes, influence of
magnetic fields is different from that for essentially dissipative
transport~\cite{comment-1}  and requires additional investigation.

{\it Pure ballistic} FETs in magnetic fields were considered in paper,~\cite{Rees}
where unusual distributions of electron concentration and velocity, and kink-like
current-voltage characteristics were predicted.
In the present paper, we analyze peculiarities of carrier transport and properties of
{\it quasi-ballistic} FETs subjected to a magnetic field. This facilitates
understanding of behavior of such FETs at actual momentum dissipation of electrons
and can be used for developing nanoscale devices for particular applications.

\section{Basic assumptions and equations}

It has long been known that,
result of action of transverse magnetic fields on electron transport
(the Hall effect) strongly depends on specimen geometry and patterns of
current/electric field.~\cite{MR-old,MR-old-2}  For a rectangular specimen
with length  $L$ along the current and width $W$ transverse to the current,
the Hall electric field/voltage builds up if $L \gg W$, then no transverse
current and magnetoresistance arise for this case. In recent paper~\cite{Kwon}
for a specially designed gated structure with $L \gg W$, the Hall voltage
was studied in detail for linear and nonlinear transport regimes.
In opposite case $L \ll W$, which is typical for FETs geometry,\cite{Zee}
the Hall field/voltage is shortened by source and drain contacts.
Instead, both transverse Hall current and magnetoresistance appear.
Below we will consider FETs of this geometry.

\begin{figure}
\includegraphics[scale=0.2]{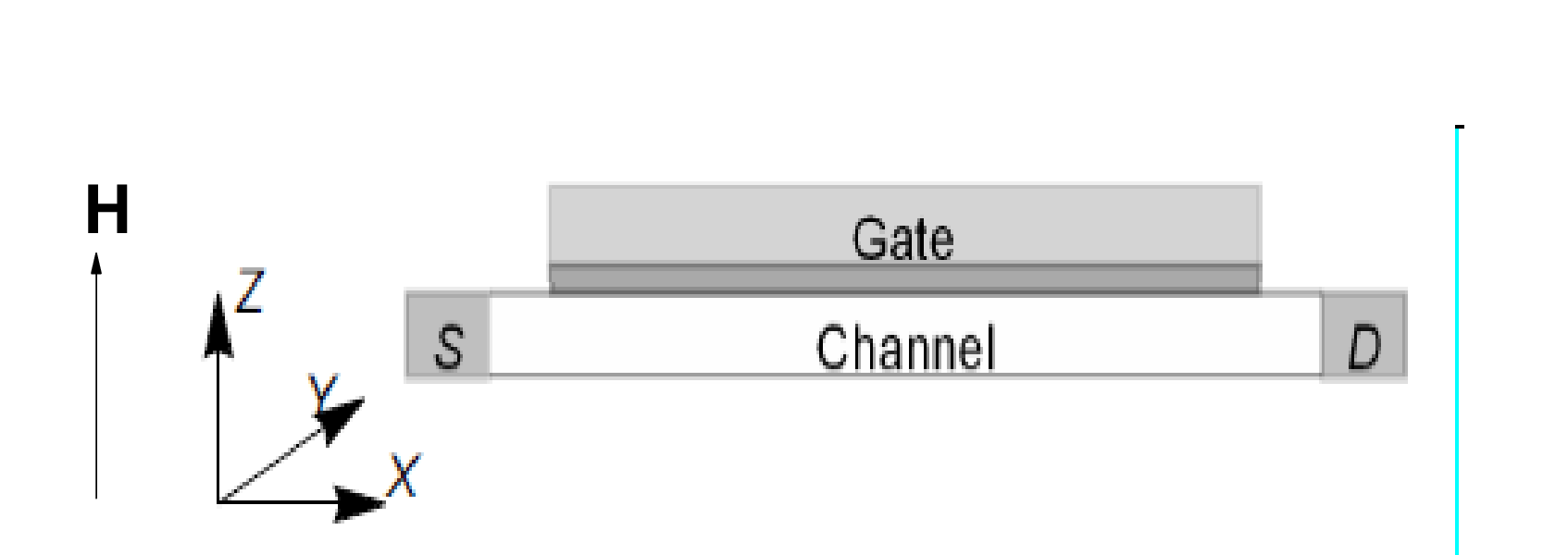}
 \caption{A sketch of FET under consideration.}
 \label{fig-1}
\end{figure}

We will use the Dyakonov-Shur model, for which  the electron transport
is considered as a charged fluid confined in a narrow layer and governed
by  hydrodynamic equations.~\cite{D-Sh-1993}
The electrons are characterized by the area density, $n$, and
by the velocity, ${\bf v}=\{v_x, v_y\}$, induced by  source-drain
electric bias, $\phi$, and a transverse magnetic field, $H_z = H$.
For the above assumed geometry of FET, we can consider that all quantities
$n,\,{\bf v},\,\phi$ are dependent only on coordinate along source-drain direction, $x$.
In the frame of the gradual channel approximation~\cite{gr-chan-appr-1,Mitin},
the local potential is supposed to be proportional to the electron density.
The complete system of equations for this model of quasi-ballistic
FET reads:
\begin{gather}
\frac{\partial v_x}{\partial t} + v_x \frac{\partial v_x}{\partial x} +\frac{v_x}{\tau}
= \frac{e}{m} \frac{\partial \phi (x,h)}{d x} - \frac{e}{c\, m} v_y H\,, \\
\frac{\partial v_y}{\partial t} + v_x \frac{\partial v_y}{\partial x} +\frac{v_y}{\tau}
=\frac{e}{c\, m} v_x H\,, \label{eq-vy} \\
\frac{\partial n}{\partial t} + \frac{\partial j_x}{\partial x} = 0\,,\,\,\,j_x = n v_x\,,
\\
\phi(x,z) = -\frac{4 \pi e}{\kappa} n(x) z + \phi_g\,\,\,\,(0 \leq z \leq h).
\end{gather}
These equations are for the frame of reference presented in Fig.~\ref{fig-1},
where geometry parameters of the FET under consideration are indicated;
the conductive layer and the gate are situated at $z= h$ and $z=0$, respectively.
The voltage applied to the gate is $\phi_g$; $m$ and $-e$ are
the electron effective mass and the
electron charge, $\kappa$ is the dielectric constant, $\tau$ is an electron relaxation time,
$c$ is the light velocity, the last terms in Eqs. (1) and (2) are projections of the Lorentz
force; $j_x$ is the electron source-drain flux density.
Eq.~(3) obtained for the gradual channel
approximation is valid at $L \gg h$ and characteristic scale of $\phi(x)$ variation is
much larger than $h$.

\section{The steady-state solutions}

For the steady-state, Eq.~(3) gives for the electron flux density: $j_x= n\,v_x=j_0$
with $j_0$ being an integration constant. It is convenient to solve the system (1)-(4) at a
given $j_0$ and then to find the voltage drop on the device, $\phi(L)$ and other
characteristics. The boundary conditions for Eq.~(1), (2) are $v_x(0) = j_0/n_s$
and $v_y(0)=0$, with $n_s=n(0)$ being the electron area density near the source.
For what follows, we introduce dimensionless variables and parameters:
\begin{gather} \label{sc-1}
\xi=\frac{x}{L}\,,\,\,V_{x,y}=\frac{v_{x,y}}{v_s}\,,\,\,N:=\frac{n}{n_s}\,,\,\,J_x =\frac{j_0}{j_{sc}}\,,\\
\Phi(\xi) = \frac{[\phi(\xi,-h)-\phi_g]}{u_{sc}}\,,\,\, \,
 u_{sc}=\frac{4 \pi e h n_s}{\kappa}\,, \label{sc-2} \\
 j_{sc} =\sqrt{\frac{4 \pi e^2 h n_s^3}{m \kappa}}\,,\,\,\,
{\cal B}= \frac{L}{\tau\,v_p}\,,\,\, v_p = \sqrt{\frac{4 \pi e^2 h n_s}{m\,\kappa}}, \label{J-B}\\
\,\,\,\Omega = \omega_c\,\tau,\,\,\,\Omega' =\omega_c\,\frac{L}{v_p},\,\,\, \omega_c = \frac{e}{c\,m} H\,. \label{Omega}
\end{gather}
Here the scaling parameters for the flux, $j_{sc}$,  and the potential, $u_{sc}$, account for the
effect of interaction of the electrons with the metal gate.
Note, in these notations, dimensionless  voltage drop on the conductive channel
is $U=\Phi(1) - \Phi (0)=N(0)-N(1) \equiv 1-N(1)$. Factor $\cal B$ is the only parameter
dependent on kinetic characteristic, which is the electron relaxation time, $\tau$;
$\omega_c$ is the cyclotron frequency. For the quasi-ballistic regime, it is convenient to
characterize the magnetic field by the dimensional cyclotron frequency, $\Omega$.
Parameter $v_p$ is known as the velocity of gated plasmons~\cite{pl-resonance-1,pl-resonance-2}; it becomes especially important parameter for near-ballistic regime with
$\tau \gg \tau_{tr},\,\,\tau_{tr} = L/v_p$, i.e., ${\cal B} \ll 1$.  For the latter regime, the characteristic time is the transit time, $\tau_{tr}$,  and relevant dimensional cyclotron frequency is $\Omega'$.

For the steady-state problem at a given electron flux/current, $J_x$,
dimensionless electron concentration, $N(\xi)$, and velocity, $V_x(\xi)$, are
related through equation $N(\xi)\,V_x(\xi)=1$. Then, accepting notations
(\ref{sc-1})-(\ref{Omega}) the dynamic equations (1), (\ref{eq-vy})
can be rewritten in terms $N(\xi),\,V_y(\xi)$:
\begin{gather} \label{N-eq}
\frac{d N}{d \xi} = \frac{{\cal B} J_x N^2}{J_x^2-N^3}\,(1 + \Omega N V_y)\,,\\
\frac{d V_y}{d \xi} =\frac{\cal B}{J_x}\,( - N V_y + \Omega)\,, \label{Vy-eq}
\end{gather}
with boundary conditions at the cathode side $N(0)=1,\,V_y(0)=0$.

Two general conclusions immediately follow from this system of equations.
First, reasonable solutions~\cite{comment-2} exist only at $J_x < 1$ (see discussion below).
Second, the first integral of equations can be found in an explicit form:
\begin{equation} \label{f-integral}
\Omega J_x^2 V_y + J_x^2 \frac{N-1}{N} + \frac{1-N^2}{2} = {\cal B} J_x (1+ \Omega^2) \xi \,.
\end{equation}
This relationship allows to find the equation of the first order for $N(\xi)$ and
facilitates determination of $V_y(\xi)$ and the Hall current density $J_y(\xi)$,
when the dependence $N(\xi)$ will be found.
\begin{figure}
\includegraphics[width=4cm,height=4cm]{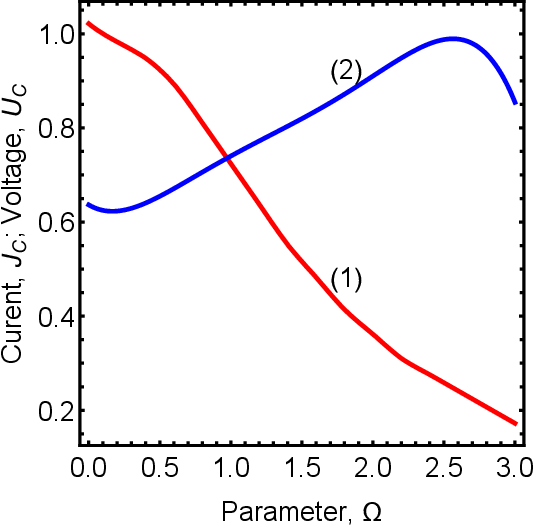}
\caption{Critical dimensional currents, $J_c$ (scaled to their values at
$\Omega=0$) and voltages, $U_c$
as functions of dimensionless magnetic field, $\Omega$.}
\label{fig-1-1}
\end{figure}
The basic Eqs.~(\ref{N-eq}), (\ref{Vy-eq}) contain a given current density $J_x$
and two parameter: ${\cal B}$, which as easy to find is dimensionless FET resistance density,
and $\Omega \propto H$. At that, the product ${\cal B}\, \Omega$ does not dependent on
the relaxation time. The case ${\cal B} \ll 1$ corresponds to nearly ballistic electron transport.
The opposite case, ${\cal B} \gg 1$, is relevant to dissipative transport.
\begin{figure}
\includegraphics[width=3cm,height=3cm]{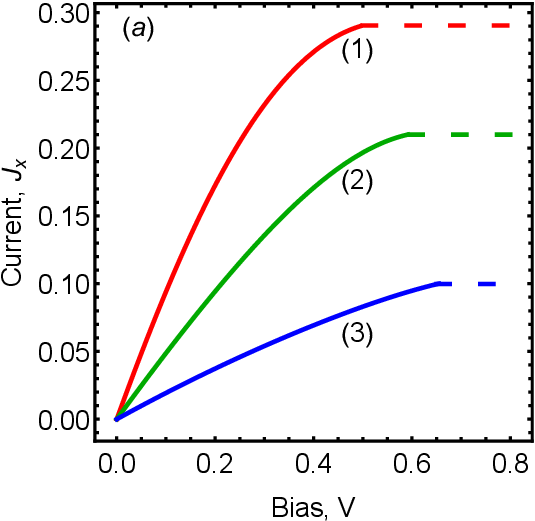}
\includegraphics[width=3cm,height=3cm]{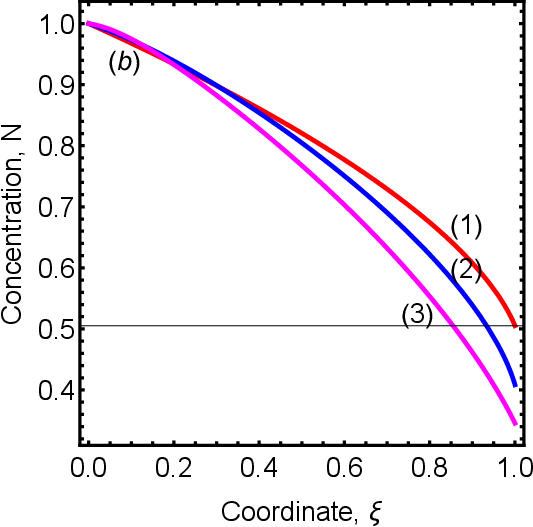}
\includegraphics[width=3cm,height=3cm]{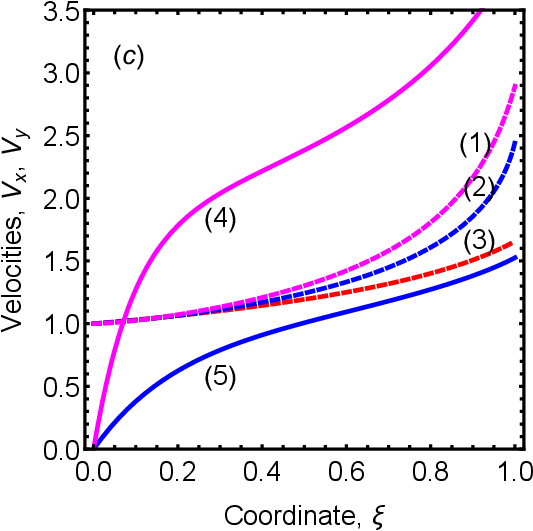}
\includegraphics[width=3cm,height=3cm]{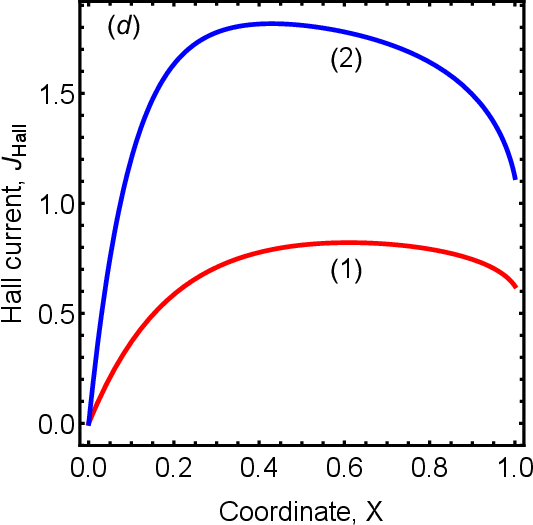}
  \caption{Results of calculations of quasi-ballistic FET in magnetic fields
  at ${\cal B}=1$. (a): The dimensionless current-voltage characteristics.
(b): Spatial distributions of the electron concentration, $N(\xi)$.
(c): Distributions of electron velocities, $V_x(\xi),\,V_y(\xi)$. (d):
Distributions of the Hall current densities, $Jy(\xi)$.
In (a), (b) and (c) curves 1, 2, 3 for $\Omega =0,\,1,\,2$,  respectively.
In (c)  curves 1 - 3 for $V_x(\xi)$; curves 4, 5 for $V_y(\xi)$ at $\Omega =1,\,2$.
In (d) curves 1, 2 for the Hall current densities.
Results in (b)-(d) are given for critical currents given in the text.}
 \label{fig-2}
\end{figure}
 The general property of the Dyakonov-Shur model and
Eqs.~(\ref{N-eq}), (\ref{Vy-eq}) is the existance of physically reasonable solutions
only in some interval of currents, $[0,J_c]$ and corresponding voltage drop interval $[0, U_c]$,
with the critical values $J_c$ and $U_c$ dependent on ${\cal B}$ and the magnetic
field magnitude, $H$. At approaching
the critical voltage, $U \rightarrow U_c$ the current has a tendency to saturate.
Dependencies of the critical values, $J_c,\,U_c$ on the magnetic field is shown in
Fig.~\ref{fig-1-1}. Note, in quasi-ballistic FETs actually there is no pinching-off
of the conductive channel. The real pinch-off effect arises at large $\cal B$ (at strongly dissipative
transport). Instead, the electric field in the channel diverges near the drain contact
for the limit $J \rightarrow J_c$.

In Fig.~\ref{fig-2} (a) we present the current-voltage characteristic for
quasi-ballistic FET with ${\cal B} =1$, i.e.., the case of moderate relaxation.
Saturation portions of $J-U$-dependencies are shown conditionally.
As seen, the magnetic field suppresses source-drain current, decreases its critical
value, $J_c$, and increases critical voltage, $U_c$: $J_c \approx 0.29, \,U_c \approx 0.6$
at $\Omega =0$ ;  $J_c \approx 0.21, \,U_c \approx 0.68$; $\Omega =1$, $J_c \approx 0.1, \,U_c \approx 0.77$  at  $\Omega =2$. Corresponding spatial distributions of the electron concentration
are given in Fig,~\ref{fig-2} (b), which evidences that magnetic field induces more
rapid decrease of the concentration along the channel. Magnetic field also leads to
an increase of the electron velocity along the channel, $V_x(\xi)$, and generates
transverse velocity, $V_y(\xi)$, which can exceed $V_x$ at high magnetic fields,
as seen from Fig.~\ref{fig-2} (c). Higher velocities of the electrons , $V_x,\,V_y$,
are achieved for the expanse of larger voltage drop in the presence of magnetic field.
The Hall current densities, $Jh=V_y(\xi) N(\xi)$, are nonuniform, they are shown
in Fig.~\ref{fig-2} (d) for $\Omega=1,\,2$. The total Hall currents,
$ I_H = \int_0^1 d \xi J_H(\xi)$,for the presented distributions are $I_H=0.67$ and
$1.56$, respectively (while total source-drain currents in the FET should be estimated as
$I_{sd} =J_x W/L$).

\begin{figure}
\includegraphics[width=3cm,height=3cm]{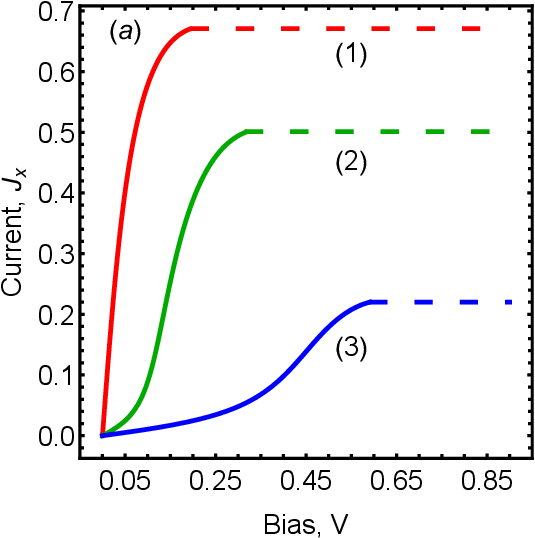}
\includegraphics[width=3cm,height=3cm]{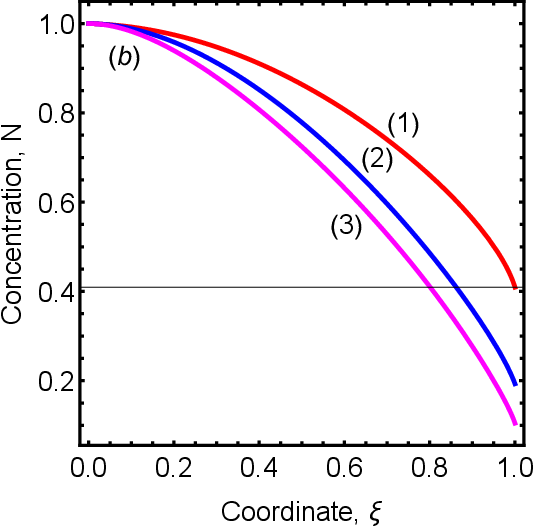}
\includegraphics[width=3cm,height=3cm]{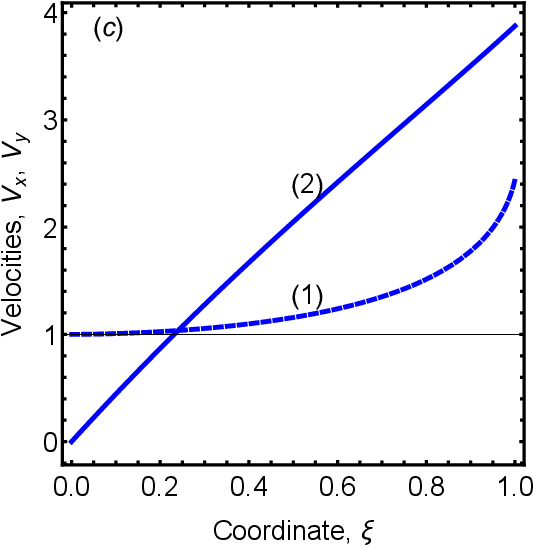}
\includegraphics[width=3cm,height=3cm]{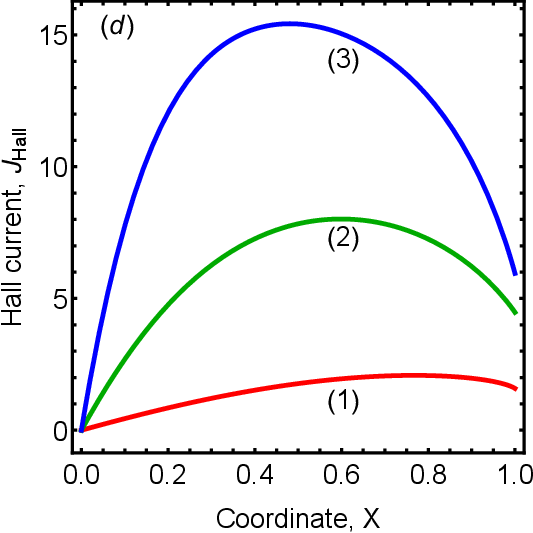}
  \caption{Results of calculations of near-ballistic FET in magnetic fields
  with ${\cal B}=0.1$. (a): The dimensionless current-voltage characteristics.
(b): Spatial distributions of the electron concentration, $N(\xi)$.
(c): Distributions of electron velocities, $V_x(\xi),\,V_y(\xi)$. (d):
Distributions of the Hall current densities, $Jy(\xi)$.
In (a), (b) curves 1 -4 for $\Omega' = 1,\,1.5,\,2$,  respectively.
In (c)  curves 1,2 for $V_x(\xi)$ at $\Omega'=1$; .
In (d) curves 1, 2, 3 for the Hall current densities at $\Omega' = 1,\,1.5,\,2$.
Results in (b), (c), (d) are given for critical currents given in the text.}
 \label{fig-bal}
\end{figure}

\section{The near-ballistic regime}

As seen from the Fig.~\ref{fig-2} (a), for quasi-ballistic transport regime
(${\cal B} \gtrsim 1$) the current-voltage characteristics have the form
typical for FETs.~\cite{Zee,gr-chan-appr-1} Situation changes cardinally for
the {\it near-ballistic} transport regime, when ${\cal B} \ll 1$.
In Fig.~\ref{fig-bal} we present the current-voltage characteristics calculated at
${\cal B}=0.1$, that corresponds to 10 times increase in the relaxation time, $\tau$,
comparing the results in Fig.~\ref{fig-2}.
 To keep amplitudes of the magnetic field the same, as above, we used the parameter
 $\Omega' = {\cal B} \Omega$, that does not dependent on $\tau$ and is more adequate
characteristic of the $H$-field for near-ballistic regime (see Eqs.~(\ref{Omega})).
One can see, that the magnetic field not only suppresses the currents, but
strongly increases the critical currents/voltages and modifies the $J-U$-dependence
toward  unusual kink-like shape with an inflection point and
two successive portions of this dependence with positive and negative curvatures.

The origin of such a behavior of the $J-U$-characteristic can be understood by considering
the {\it  pure ballistic} limit of the electron transport. First, we should mention
that in the absence of magnetic field the Dyakonov-Shur model has no solutions for this limiting regime. For a finite value
of the magnetic field, i.e., finite $\Omega$, setting ${\cal B}=0$ in
Eqs.~(\ref{N-eq}) - (\ref{f-integral}), we find following implicit expression
for $N(\xi)$ at a given $J$:
\begin{equation} \label{case-B=0}
1-N(\xi) -J^2 \frac{1-N(\xi)^2}{2 N(\xi)^2} =\frac{1}{2} \Omega'^2 \xi^2\,.
\end{equation}
Using this equation for $\xi=1$, we find that the electron transport arises at finite threshold voltage
$U_{th} = \Omega'^2/2 <1 $ with small, square root dependent  current above the threshold:
$$
J = \sqrt{\frac{2\,(U-U_{th})} {U_{tr} \,(2-U_{tr})}}\,(1-U_{tr})\,.
$$

\begin{figure}
\includegraphics[width=6cm,height=6cm]{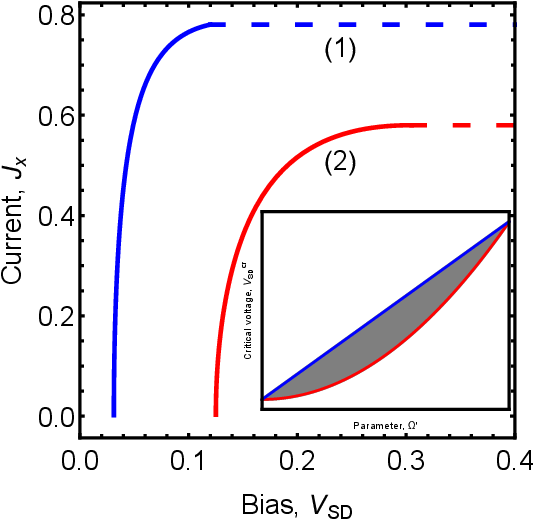}
\caption{Pure ballistic FET in magnetic field. Main panel - $J-U$-characteristics for
$\Omega'=0.3$ and $0.54$.
Inset: region of existing solutions of Eq.~(\ref{case-B=0}) in the
$\{U, \Omega'\}$-plane.}
\label{fig-5}
\end{figure}

Existence of a threshold for the current, absence of a liner portion of $J-U$-characteristic
and its negative curvature
explain the kink-like behavior of this dependence. Arising strongly nonlinear
resistance of the device is not related to conventional dissipative processes.
Instead, it arises due to pump over electric energy into the Hall currents.
Indeed, calculations indicate that the Hall velocity increases much faster than the
velocity in the source-to-drain direction.
From Eq.~(\ref{case-B=0}) it follows that the source-drain currents
occur in a finite over-threshold  interval of the voltage, $U$, with a tendency
of suppression of the current values, when  the magnetic field  increases. At high magnetic
fields ($\Omega' > \sqrt{2}$),  solutions of Eq.~(\ref{case-B=0}) and the currents are absent.
In the inset to Fig.~\ref{fig-5}, the region of existing solutions
to Eq.~(\ref{case-B=0})  is shown  in the $\{U,\,\Omega' \}$-plane. In the main panel of this figure, examples of $J-U$-characteristics for the pure ballistic regime are presented.

Consideration of even small electron dissipation restores the currents
in the sub-threshold voltage range, as demonstrated in Fig.~\ref{fig-bal} (a).
Results in Fig~\ref{fig-bal} (b) show that in the near-ballistic regime
the concentration decreases faster along the channel and the total voltage drop
are larger comparing to the quasi-ballistic regime (see Fig.~\ref{fig-2} a).
This, particularly, corresponds to larger electric fields acting the electrons.
Faster growth of the Hall velocity, $V_y$, in comparison with the velocity $V_x$ is
illustrated in Fig.~\ref{fig-bal} (c)). This explains high values of the Hall current
densities, as demonstrated in Fig.~\ref{fig-bal} (d). Respective total Hall currents
are $I_H= 0.57, 2.39$ and $ 5.6$ at  $\Omega'=1, 1.5$ and $2$, respectively.

\section{Magnetoresistance at arbitrary currents}

Dimensionless FET resistance in nonlinear current regime can be defined as
\begin{equation} \label{MR-1}
R=\frac{U}{J} = \frac{1-N(1)}{J}\,.
\end{equation}
Thus, dependence of the electron concentration at the drain side, N(1), on magnetic field
directly gives the magnetic resistance of the FET. Note, in the absence of the $H$-field
for the linear regime (small $U$ and $J$) $R = R_0 ={\cal B}$.

In the case of small magnetic field, the magnetoresistance for arbitrary currents
can be calculated as follows.
Considering in Eqs.~(\ref{N-eq}), (\ref{Vy-eq}) terms proportional to $\Omega$ as
small, we find the solution of Eq.~(\ref{Vy-eq}) in the form
\begin{equation} \label{MR-2}
V_y(\xi) = \frac{{\cal B}\, \Omega}{J_0}
\int_0^{\xi} d \xi'e^{\frac{{\cal B}}{J_0} [\int^{\xi'} d x' N_0(x')-
\int^{\xi} d x N_0(x)]}\,,
\end{equation}
where $J_0 \equiv J_x$ and $N_0(\xi)$ are the current and electron concentration in the absence
of the $H$-field. From Eq.~(\ref{N-eq}),  we obtain perturbation of the electron
concentration, $N_1(\xi)$:
\begin{gather} \nonumber
N_1(\xi)=\frac{{\cal B}^2\, \Omega^2 (N_0(\xi))^2}{J_0^2- (N_0(\xi))^2} \times \\
\int_0^\xi d \xi' N_0(\xi') \int_0^{\xi'} d\xi''e^{\frac{{\cal B}}{J_0} [\int^{\xi''} d x' N_0(x')- \int^{\xi'} d x N_0(x)]}\,.\label{MR-3}
\end{gather}

Dependence $N_0(\xi)$ in the implicit form is given by the relationship
(\ref{f-integral}) at $\Omega=0$. Eq.~(\ref{MR-3}) shows that always $N_1(\xi) < 0$ and
proportional to $\Omega^2$. Thus, for small magnetic fields the FET magnetoresistance,
$\Delta R = - N_1(1)/J_0$, is positive and quadratic in the $H$ field.

\begin{figure}
\includegraphics[width=3cm,height=3cm]{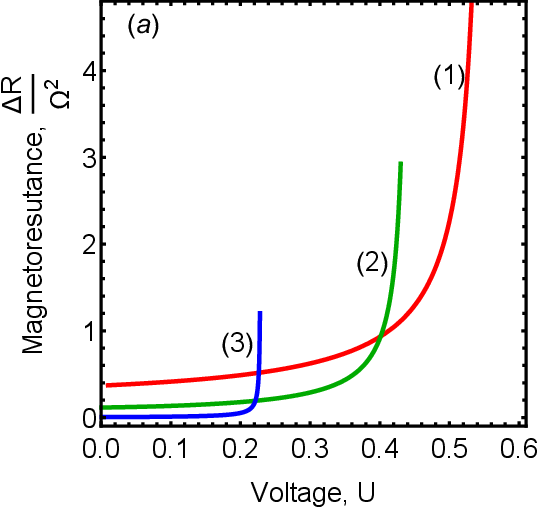}
\includegraphics[width=3cm,height=3cm]{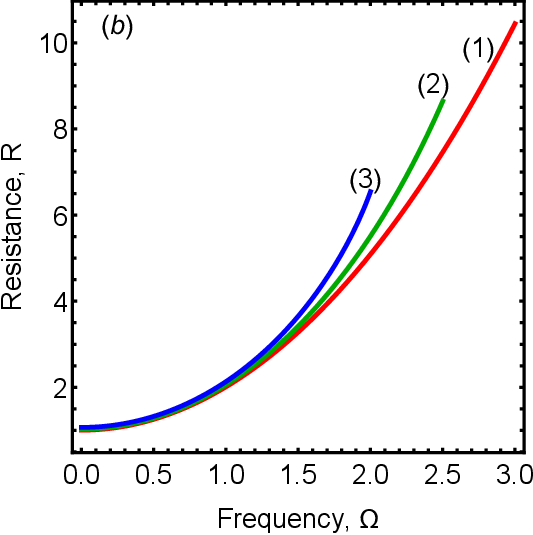}
\includegraphics[width=3cm,height=3cm]{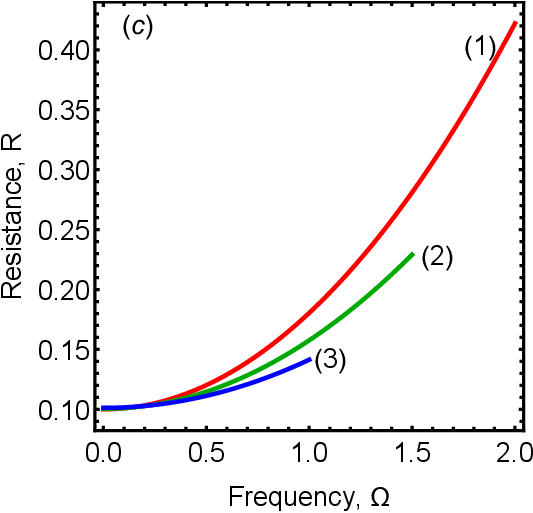}
\caption{(a): Factor $\Delta R/\Omega^2$ defining low H-field magnetoresistance according
to Eq.~(\ref{A1}); curves 1 - 4 are for ${\cal B} = 2, 1, 0.5, 0.1$, respectively.
(b), (c): Total resistance of FET as function on $\Omega$ for different values of current.
(b): ${\cal B}=1$, 1 - J=0.01, 2- 0.05, 3- 0.1, 4 - 0.2, 5 - 0.29.
(c): ${\cal B} =0.1$, 1 -  J=0.021, 2 - 0.224, 3 - 0.579  }
\label{fig-4}
\end{figure}

Eq.~(\ref{MR-3}) can be rewritten to the form dependent only of the voltage drop, $U_0$
and respective current $J_0$.
Indeed, from Eq.~(\ref{N-eq}) at $\Omega=0$ it follows, that
$$
d \xi = \frac{J_0^2-(N_0(\xi))^3}{{\cal B} J_0 N_0^2(\xi)} d N_0\,.
$$
Thus in Eq.~(\ref{MR-3}) we can change integrations over $\xi',\,\xi''\,$ by integrations
over $N',\,N''$. The result gives the following closed form for the magnetoresistance:
\begin{gather} \nonumber
\Delta R = \Omega^2 \frac{(1- U_0)^2}{J_0^2[(1-U_0)^3-J_0^2]} \,
 \int_1\limits^{(1-U_0)} \frac{d N}{N} [J_0^2-N^3]\times \\
\int_1^N \frac{d N'}{N'^2}[J_0^2-N'^3]\, e^{[P(N')-P(N)]} \label{A1}
\end{gather}
with
$$
P(N) = \frac{\cal B}{J_0} \int_1^N \frac{d N'}{N'^2}[J_0^2-N'^3]\,.
$$
Eq.~(\ref{A1}) allows to find the magnetoresistance at small magnetic fields
for both linear and nonlinear current/voltage regimes. In Fig.~\ref{fig-4} (a) we
present the factor $\Delta R/\Omega^2$ for different values of dimensionless FET
resistance at small biases, $\cal B$. One can see that the magnetoresistance
increases in the nonlinear current regime with the tendency to divergence approaching
 critical currents/biases. Furthermore, while at low bias the   magnetoresistance
is larger for more dissipative channels ($\cal B$ is larger),  for
the nonlinear regime, in opposite, this resistance is larger for channels
with larger ballisticity ($\cal B$ is smaller).

At large magnetic fields, the FET resistance is presented in Figs.~\ref{fig-4} (b), (c)
as a function of $\Omega$ for different values of the current. The resistance
was calculated as follows. For a  given $\Omega$, we find the $J-U$-characteristic.
Then, at a fixed $J <J_c$ the voltage drop $U$ can be determined and $R$ is calculated according
to Eq.~(\ref{MR-1}). In panels (b) and (c) the results are given for quaisi-ballisic
(${\cal B} =1$) and  near-ballistic (${\cal B}=0.1$) FETs, respectively.
For the latter case the magnetoresistance  is considerably larger despite smaller
electron dissipation. When current changes, sequences of the curves $R(\Omega)$
are opposite for these two transport regimes, as indicated in Figs.~\ref{fig-4} (b), (c).

\section{Numerical estimates}

The above resilts presented in the dimensional form can be applied to analyze
quasi-ballistic FETs made of different materials and with different geometry parameters.
For that the scaling parameters of voltage, $u_{sc}$, current, $j_{sc}$,
velocity, $v_p$, as well as  parameter ${\cal B}$, have to be estimated.
 The scaling parameters are determined by the electron characteristics, charge, mass,
area concentration, and dielectric surrounding and distance between the channel and the gate, $h$.
These parameters reflect the essentiality of electron-metal gate interaction.
The parameter ${\cal B}$ also depends on the electro relaxation time, $\tau$, and the channel length $L$.

For  numerical estimates, we consider the FET with
the GaAs conductive channel. We set $m= 0.063\,m_0\,,\kappa =10.9\,,
n_s =10^{12}\,cm^{-2}$ and $h = 5 \times 10^{-7}\,cm$. Then, the scaling parameters
for voltage, current density and velocity are $u_{sc} = 83\,mV$, $e j_{sc} =7.3\,A/cm$
and $v_p =4.6 \times 10^7 cm/s$. We consider properties of the FET at
two temperatures: $77\,K$ and $300\,K$.
For $77\,K$ at the mobility $\mu = 3 \times 10^5\,cm^2/V s$,
we find $\tau \approx 1.2 \times 10^{-11}\,s$ and $B=1$ at $L_x \approx 5.4\,\mu$.
At that the dimensionless cyclotron frequency $\Omega =1$ corresponds to
the magnetic field $H = 0.04\,T$. The, numerical results presented in Fig.~\ref{fig-2}
are relevant to these estimates.

For $300\,K$ and $\mu = 800\,cm^2/V s$, we find $\tau \approx 3 \times 10^{-13}\,s$,
$\Omega=1$ corresponds to the magnetic field $H = 1.3\,T$. Parameter ${\cal B}=1$ is
realized at $L_x= 0.14\,\mu m$. Thus, the results  of Fig.~\ref{fig-2} are consistent
with these estimates for quasi-ballistic FET at room temperature.

The above estimates show that the near-ballistic regime with characteristics presented in
Fig.~\ref{fig-bal} can be realized in GaAs FET with $L_x=0.54\,\mu m$ for nitrogen
temperature, when ${\cal B} =0.1$ and $\Omega'$ corresponds to $H=0.04\,T$.
Similar estimates we found for other III-V compounds.

\section{conclusions}

It is known that influence of transverse magnetic field on properties of
FET is significant at particular device geometry, when width of the
active channel, $W$, is much large than its source-to-drain length, $L$,
and Hall field/voltage is shortened by source and drain contacts,
instead Hall currents arise (this can be called quasi-Carbino case). For {\it the
dissipative  regime} of electron transport in the channel, the magnetic field
effects have been studied in a number of papers (see,
for example~\cite{H-field-1,H-field-2,H-field-3}).
These studies and their conclusions
are correct when electron scattering time is much less than their transit time
through the device and, particularly, the drift-diffusion approach~\cite{Mitin} can be applied.
In the present paper we analyzed magnetic field effects in FETs with quasi-ballistic and
near-ballistic electron transport described by hydrodynamic equations
of the Dyakonov-Shur model.~\cite{D-Sh-1993}

For the considered quasi-ballistic FETs, the main effects of the magnetic field
are: significant suppression of currents, increase of saturation voltage, induction of large
Hall currents, etc.
 The density of the Hall currents can exceed densities of the source-drain currents.
We found that in quasi-ballistic FETs ($\cal B$ is finite) there is no real
pinch-off in the channel. Instead, the electric field diverges in the drain
side of the device  just before the saturation regime.

  It t is pertinent to note that for the considered geometry of FETs,
uniformity of both the equipotential lines and  Hall currents in the transverse direction
breaks down near the edges of the channel ($y= \pm W/2$), where the current lines exit/enter
the source/drain contacts. Thus, to observe predicted above $J-U$-characteristics
the following condition for the total source-drain current, $I_{sd}$, should be fulfilled: $I_{sd}(U) = J(U)\,W \gg I_H (U)$ with $I_H$
been the total Hall current estimated above.

Suppression of currents trough the device means increasing resistance. For small magnetic
fields, we found general expression for the classical magnetoresistance
(proportional to square of the $H$ field).This expression revealed unexpected
result: in nonlinear current regime the magnetoresistance is larger for the
devices with higher ballisticity. This property is  confirmed
by the analysis at high magnetic fields, when the device
resistance can raise several times depending on the current.

For near-ballistic transport regime in FETs, the magnetic field cardinally modifies
the $J-U$-characteristics toward to kink-like behavior. Analysis of such a regime,
including the pure-ballistic case, shows that this behavior and strongly nonlinear
resistance at almost absent dissipation arise due to pump over electric energy
into the Hall currents. We determined regions of voltages, currents and magnetic
fields, where these unusual effects occur.

Numerical estimates show that for FETs channels of high mobilities (e.g., low
temperature, GaAs)
significant effects can be observed in moderate magnetic fields, $ \sim 0.1\,T$.
At that, quasi-ballistic regime occurs for a few microns channel length, while
for the submicron length the near-ballistic regime can be realized.
At moderate mobilities (e.g., room temperature, GaAs), only quasi-ballistic regime can occur
and  the large magnetic induced effects require the fields $\sim 1\,T$.

Summarising, we studied semiclassical quasi-ballistic electron transport in short
FETs subjected to magnetic field. We determined spatial distributions of
electron concentrations, velocities, Hall currents and voltages along
the channels. These distributions can be measured by using different
scanning probe microscopy techniques. Large Hall currents in the FETs
should induce magneto-optical effects, which
can be studied in THz spectral range for a single device, as well as for transistor
 array, when the response of the devices is synchronized by the external
signal.~\cite{FET-array} We defined basic properties of
current-voltage characteristics of the FETs in magnetic
field, among these we have analyzed kink-like $J-U$-characteristic
of the near-ballistic device. Peculiarities of magnetoresistance of
quasi-ballistic FETs are studied for low and high magnetic fields,
and different current regimes.
We suggest, that the found results contribute
to the physics of short FETs and can be used for developing
nanoscale devices for particular applications, including THz
emitters/detectors,~\cite{H-field-2,THz-det-1,THz-det-2,THz-gener-1}
magnetic micro sensors, etc.

\begin{acknowledgments}
This work was conducted in frame of the long-term program of support of
the Ukrainian research teams at the Polish Academy of Sciences,
project LTP-NAS-KOCHELAP-A7-11-0008.
\end{acknowledgments}

\section*{Author Declarations}

Conflict of Interest
The authors have no conflicts to disclose.

\section*{DATA AVAILABILITY}
The data that support the findings of this study are available
from the corresponding author upon reasonable request.


\begin{references}
\bibitem{comment-1}
For example, characterization of electron kinetics by a mobility is
not correct for quasi-ballistic transport. Indeed, mobility can
be introduced only when scales of spatial
nonuniformity well exceeding the mean free path, which is not discussed case.
Otherwise, mobility being a local characteristic will be found dependent
on device sizes (for example, on the gate length, as in papers~\cite{H-field-1,H-field-2,H-field-3}).
\end{references}
\end{document}